\documentclass[sigconf]{acmart}
\AtBeginDocument{%
  }

\setcopyright{none}

\acmDOI{10.48550/arXiv.2508.05223}
\acmConference[LIMITS '25]{11th Workshop on Computing within Limits}{June 25--26, 2025}{Online}

\acmISBN{}

\usepackage[textsize=footnotesize, textwidth=15mm]{todonotes}
\presetkeys{todonotes}{fancyline}{}
\usepackage[inline]{enumitem}
\usepackage[english=american,babel=false]{csquotes}
\usepackage{url}
\usepackage{hyperref}

\begin{document}

\title{Resistance Technologies: Moving Beyond Alternative Designs}

\author{Iness Ben Guirat}

\email{iness.ben.guirat@ulb.be}
\orcid{0000-0002-8766-594X}

\affiliation{%
  \institution{Universit\'e Libre de Bruxelles}
  \city{Brussels}
  \country{Belgium}
}

\author{Jan Tobias M\"uhlberg}

\email{jan.tobias.muehlberg@ulb.be}
\orcid{0000-0001-5035-0576}

\affiliation{%
	\institution{Universit\'e Libre de Bruxelles}
	\city{Brussels}
	\country{Belgium}
}
\settopmatter{printfolios=true}
\pagenumbering{arabic}
\renewcommand{\shortauthors}{Ben Guirat \& M\"uhlberg}

\begin{abstract}
The discourse about sustainable technology has emerged from the
acknowledgment of the environmental collapse we are facing. In this paper,
we argue that addressing this crisis requires more than the development of
sustainable alternatives to current online services or the optimization of
resources using various dashboards and AI. Rather, the focus must shift
toward designing technologies that protect us from the consequences of the
environmental damages. Among these consequences, wars, genocide and new
forms of colonialism are perhaps the most significant. We identify
\enquote{protection} not in terms of military defense as Western States
like to argue, but as part of sovereignty. We seek to define the term of
\emph{Resistance Technologies} for such technologies, arguing further that
anti-surveillance technologies are a foundational component of sovereignty
and must be part of future conversations around sustainability.  Finally,
our paper seeks to open a discourse with the Computing-within-Limits
community and beyond, towards defining other essential aspects or concepts
of technologies that we see as core values of \emph{Resistance
Technology}.
\end{abstract}


\maketitle

\section{Introduction}

Humanity has already crossed at least seven out of nine planetary
boundaries~\cite{persson_outside_2022,findlay_ocean_2025},
with profound consequences for all life on Earth~\cite{poertner2022climate}. The impacts of
this transgression, particularly regarding climate change, are no longer
abstract. They are unfolding now in the form of extreme weather events,
melting ice caps, and heat waves fueling catastrophic wildfires. In
response, academics and activists are increasingly turning to
technology---not only to analyze these crises but to design systems that
might help us avoid further devastation. These efforts range from
theoretical frameworks like Just Sustainability and Justice Designs to more
practical approaches, such as sustainable software alternatives and
dashboards for optimizing resource use. While these developments are
promising, we are still far from achieving sustainable computing, as noted
by Christoph Becker in Insolvent~\cite{becker2023insolvent}.

We share the view of Eriksson et al.~\cite{eriksson2018meeting}, who write
that \enquote{it is hard to argue that the future of computing will lead to
something that is different from business-as-usual.} This is due to several reasons. First, most attempts at producing sustainable software are forced to fit the market logic---and in doing so, they are ultimately undermined by those very market. Sustainable systems cannot compete with profit-driven models that thrive on labor and resource exploitation. Second, even when they succeed, they often face \enquote{death by acquisition,} where they are bought out, shut down, and their innovations are buried. Additionally, as Cory Doctorow argues, \enquote{enshittification}---the process by which platforms degrade over time as monetization takes precedence---is an inevitable outcome of systems designed around generating profit. The venture capital-driven ecosystem has stifled innovation by eliminating space for small businesses and concentrating wealth and power in the hands of a few dominant organizations~\cite{corydoctorow2023}. Therefore attempting to create alternative software solutions becomes largely futile, a misdirection of energy and effort. This is not to say, however, that all efforts to limit human environmental impact are useless; rather, they are insufficient on their own. As Christoph Becker has noted, it remains unclear whether such alternatives genuinely lead to sustainability, especially considering the potential for rebound effects in consumption~\cite{becker2023insolvent}.

Rather, the central objective of the sustainability community and our work should be to propose radical new designs that will prepare us for
the consequences of the ecological and climate polycrisis. Arturo Escobar refers to design as \enquote{much more than the creation of objects (toasters, chairs, digital devices), famous buildings, functional social services, or ecologically minded production. What the notion of design signals is diverse forms of life}~\cite{escobar2018designs}. Building on this, we argue that systems should be designed with careful consideration of the consequences of climate change and in anticipation of the future modes of living it will necessitate.
We acknowledge that our perspectives are rooted in a western understanding of
economics, western perspectives of sustainability and specifically of
scarcity~\cite{tahvonen2000economic}. We believe that, throughout our
reflections, we develop a vision of digital systems that emphasized
interconnectedness of humans and nature for communal well-being
and preparedness that bear strong links with notions of stewardship
(cf.~\cite{chapin2010ecosystem,mathevet2018concept}) and
earlier non-western understandings of sustainability
(cf.~\cite{throsby2016sustainability}).

The polycrisis is defined by human activity exceeding limits
beyond which our biosphere would no longer be able to
self-regulate. In consequence, the Earth system would leave the stability
of the Holocene and become increasingly inhospitable for humans.
Limits as well as human impact relating to, e.g., climate change, ocean
acidification, disruption of biogeochemical flows, land and water use, and
biosphere integrity, are studied and
quantified~\cite{persson_outside_2022}. Political bodies generally
acknowledge that, e.g., \enquote{the consumption of an average [European
Union] citizen
is outside the safe operating space for
humanity}~\cite{sala_consumption_2019}. Even knock-on effects of
unmitigated impacts are being studied and predictions assert, for example,
that \enquote{at higher global warming levels, impacts of weather and climate
extremes, particularly drought, by increasing vulnerability will
increasingly affect violent intrastate conflict}~\cite{poertner2022climate}.

So what kind of lifestyle should we imagine in the face of climate
collapse? How do we frame this lifestyle and the technological choices that
come with it? We believe that we can present sustainable and resilient technological choices
from a perspective of sufficiency or even abundance.
Critical design approaches are valuable because they ask the
\emph{why} and the \emph{how}. However they often fall short in addressing the
\emph{what}. It is not however to undermine their work, creating new
approaches to design is absolutely essential in the future designs however the
problem is hard. As Eriksson et al. remind us, imagining radically different futures is inherently difficult because our visions are constrained by the
systems we inhabit~\cite{eriksson2018meeting}.

In this paper, we propose a category for future design which we call
\textit{Resistance Technologies}: systems that are not simple alternatives but designed to be useful in the times of climate crisis, war, colonialism, growing inequalities and power imbalance, and resource scarcity. We argue that privacy is one of the core values of resistance
technologies---because it enables autonomy, dignity, and survival in a world
increasingly driven by automation, extraction, and control.  We do not
claim to have solved the challenge of future design. Rather, we attempt to
identify a common denominator across the crises we face.  We then turn to the
Computing with Limits community for discussions on other core values that
should constitute resistance technologies.

We frame our argument for the urgent need for building resistance technologies around three key questions: Why, How, and What. First, Why do we need radical design rather than mere alternatives? This section explores the necessity of moving beyond alternative solutions in times of climate crisis. Second, How can we design such systems? Here, we explore critical design approaches and their limitations within the scope of this paper. Finally, What should we design? We argue for framing future systems as resistance technologies and propose that privacy-enhancing technologies should be a core aspect for resistance technologies. We conclude by outlining a set of guiding features and questions that can support the design of future systems, and we turn to the \textit{Computing within Limits} community to further refine and expand this perspective.

\section{Why: Challenges of the Digital in a Polycrisis}

In this section, we argue for a shift away from building \enquote{green} or
\enquote{efficient} alternatives, and instead call for the development of resistant technologies in the face of the climate crisis.

As outlined in the introduction, efforts to build alternative solutions often fail---not because of technical limitations, but because they cannot compete with capitalist systems that prioritize cheaper, marketable options built on labor and nature exploitation.
More importantly, focusing solely on optimizing energy consumption or
material use can paradoxically produce unsustainable results. Tainter
argues that as societies develop solutions, they tend to introduce
additional complexity, which in turn requires more resources and can lead
to diminishing returns and increased
unsustainability~\cite{tainter_framework_2003}. M\"uhlberg expands on this
in~\cite{muehlberg_2022_sustaining_sec}, using safety and security as a
case study. He argues that integrating these features into
\enquote{sustainable} technologies often leads to increased development,
deployment, and maintenance costs. This concern aligns closely with the
concept of \enquote{Data Refusal} by Zong et al.~\cite{zong_data_2024}, which urges the sustainability community to critically evaluate and, in some cases, reject certain technological developments. Zong et al. propose four dimensions of \textit{refusal}—autonomy, time, power, and cost—to analyze practices of refusal, showing that it can be both reactive and proactive. Their framework also highlights how refusal can challenge data collection practices by governments and corporations.
The goal, of course, is not to promote unsafe or insecure software, but to question how we can achieve safety, security, and sustainability from the outset. For example reducing data collection is a far more effective strategy in terms of security and safety than continuously patching cryptographic systems which at the same time allow us to shut down entire data centers~\cite{muehlberg_2022_sustaining_sec}.
As outlined in the recent German Advisory Council on Global Change (WBGU) report on security and sustainability \cite{noauthor_security_2025}, the global and national security is increasingly shaped by environmental crises; climate change, biodiversity loss, and pollution are no longer isolated ecological issues but foundational threats to human well-being, economic stability, and geopolitical order. Therefore the WBGU argues that conventional security approaches are inadequate; instead, it calls for a holistic approach that embed environmental intelligence, prioritize ecological restoration, ensure access to essential resources and services, and address long-term inequalities across and within societies. Such an approach is essential to protect people and societies in an increasingly unstable and interconnected world.
We further explore this argument through three core challenges to sustainability:
\begin{enumerate*}[label=(\roman*)]
  \item digital colonialism,
  \item gender inequalities, and
  \item armed conflict and migration.
\end{enumerate*}

\subsection{Digital colonialism}

The control of critical digital infrastructures, means of communication such
as cables and satellites, as well as control over data and
computational resources establishes and substantiates
power~\cite{avila2020against}. As Michael Kwet has described: \enquote{this
structural form of domination is exercised through the centralized
ownership and control of the three core pillars of the digital ecosystem:
software, hardware, and network connectivity, which vests the United States
with immense political, economic, and social power. As such, GAFAM (Google/
Alphabet, Amazon, Facebook, Apple, and Microsoft) and other corporate
giants, as well as state intelligence agencies like the National Security
Agency (NSA), are the new imperialists in the international community.
Assimilation into the tech products, models, and ideologies of foreign
powers, led by the United States, constitutes a twenty-first century form
of colonization.} China has decided to oppose this US imperialism by
competing in producing software and hardware that is equally good and
perhaps cheaper. However China remains the exception to the rule in the
Global South. Except for china, no country in the Global South is able to
compete in producing technologies that are not controlled by the USA. Avila
notes three elements that disable global south countries:
\begin{enumerate*}[label=(\roman*)]
  \item resources (cables, servers and data, but also intellectual resources are being halted due to brain drain);
  \item legal architecture prevents
small countries from adopting policies that favor the production and
purchase of goods and services produced domestically, with the threat of
legal proceedings in international courts for adopting anti-competitive
measures;
  \item availability of financial capital to do research and
experiment on new designs that can compete with the well-funded products
available in the global market.
\end{enumerate*}

However, besides the fact that competing in producing software and
hardware that can compete with the US companies is basically impossible for
the majority of the global south countries, we argue that it is also not
the best strategy in the long term either.  Avila
in~\cite{avila2020against} highlights the importance of education in
resisting digital colonialism, by emphasizing the importance of students
not only learning how to be users but also able to develop their own tools.
However, we argue without a framework that describes what should we create
we are bound to create the same tools and this at best will replace the
same tools with its own flaws (non sustainable) at worst won't stand the
competition.

The global south countries are being threaten to face the harshest form on
environmental consequences therefore the technology that should be produced
in these countries should reflect the values and needs of these
communities. The technology that should be produced should be enable
countries to be self sovereign as argued by Nick
Meynen~\cite{meynen2019frontlines}.
In doing so, one can imagine a state where countries do not have to be
involved in wars over resources for example \enquote{mineral extraction} because
they developed technologies that allow them to live sustainably and they
do not need minerals to start with.

Digital colonialism is closely intertwined with surveillance, as argued by
Andr\'es Arauz in The Data of Money~\cite{datamoney19arauz}. SWIFT---the Society for Worldwide Interbank Financial Telecommunication, which controls any financial transaction between individuals and countries---is majority-owned by U.S. banks. Building anti-surveillance technologies in the Global South is therefore crucial to resisting digital colonialism.

Europe offers a telling example. In discussions around security and war,
the instinct is often to invest in more powerful military technologies. But
even when successful, this approach merely redistributes violence---it does
not prevent it. What if, instead, we designed technologies that reduce the
likelihood of war by empowering people, preserving the planet, and
fostering resilience?

\subsection{Armed Conflict, Migration, and Technology}

We have already pointed to the correlation between increasing risk of
armed conflict with increasing climate extremes such as drought, as
summarized in~\cite{poertner2022climate}. Climate change and loss of
biosphere integrity exacerbates the causes of violent conflict by making,
for example the scarcity of basic resources including food and water, much
more likely.  Burke et al.~\cite{burke_climate_2015} study the correlation
between climate and conflict and conclude that every 1-standard-deviation
increase in temperature increases the risk of intergroup conflict by
11.3\,\%. The extreme heat waves, floods and droughts that we currently
experience are already 3-$\sigma$ and 4-$\sigma$
events~\cite{hansen_perception_2012, robinson_increasing_2021}. With
increased conflict and resource scarcity also comes increased displacement
and migration. \enquote{By 2030, about 250 million people may experience
high water stress in Africa, with up to 700 million people displaced as a
result,} stipulate~\cite{poertner2022climate}. And while these displaced
people predominantly come from regions and communities that have
historically been colonized and that now suffer severe environmental
consequences of the economic activity of the Global North, we see western governments reacting with increased militarization and increased
technological effort towards managing and deterring
migration~\cite{molnar_2020_tech_testing_grounds}, rather with than
humanitarian support.

Parkinson et. al estimates global military carbon footprint  to
5.5\%~\cite{parkinson2022estimating}. This is due to the fact that the
military activities are dependent on extreme exploitation of global natural
and energy resources. If the
global military were a country, this would place it fourth in terms of its
emissions\footnote{https://www.theguardian.com/commentisfree/2024/jan/09/emission-from-war-military-gaza-ukraine-climate-change}.
Additionally, the impact of the military complex is not just carbon
footprint, but also the degradation of land, water, and the biosphere as a
whole.  This is a paradox, because on one hand the environmental damage is
done by the military complex on the other hand the environmental collapse
will cause more conflicts, wars and genocide over resources as they are becoming scarce. Governments argue that the increasing investment in
military technology is a defense against future threats yet, the investment
in this \enquote{defensive} technology exacerbates climate impacts and
thereby lead to violence.

\subsection{Feminism and Technology}

Feminist literature have argued that technology is socially constructed, shaped by dominant power structures and historical roles that often privilege masculine perspectives. Several initiatives have therefore focused on getting more women into STEM, through initiatives like \enquote{Girl Boss,} \enquote{She Codes,} and similar efforts. The underlying idea is that if more women learn to code, they will no longer be left behind by technological progress, and this will help reduce broader social inequalities.

It is important to note however that the problem behind the lack of women in tech is not just about representation. It is about the impact of this exclusion on the technologies we build. The Apple AirTag is one example: while designed to help people find their keys or bags, it has been quickly repurposed as a tool for stalking and harassment, particularly targeting women. However, clearly it is not enough for women to enter a system---the system itself must transform. As argued in ~\cite{snick2011towards}, real change demands a re-imagining of technology that supports equality, sustainability, and democratic governance.
Judy Wajcman calls for a more inclusive, feminist approach to innovation, one that integrates diverse experiences and asks critical questions about who designs technology, for whom, and based on what assumptions~\cite{wajcman2004technofeminism}.
Another growing body of literature in feminist literature is tech-related abuse~~\cite{harris2023technology}. Domestic violence is far more prevalent than we think, yet austerity measures leave women's shelters and safe spaces underfunded. In Belgium, for example, 20\% of women have experienced physical violence, with the [European Union] average at around 18\%~\footnote{\url{https://eige.europa.eu/gender-based-violence/countries/belgium?language_content_entity=en}}. Feminist scholars argue that we must include \enquote{abusive partners} as a threat model in any security or privacy system design. Government responses have often pushed for widespread CCTV deployment under the guise of increasing women's safety. However as shown by Lesli Kern in her book \enquote{Feminist City Claiming Space in a Man-Made World}~\cite{kern2021feminist} these systems do little to actually improve safety and often just displace the problem to more vulnerable neighborhoods and communities.

Female health apps are on the rise, aiming to address long-standing biases in the medical field regarding women's health. By providing designs that help women better understand their bodies, these technologies seek to fill critical gaps in care. However, growing concerns have emerged about how these apps collect, store, and potentially exploit sensitive personal data~\cite{hassan2023your}.

Once again, this highlights the critical need for privacy-enhancing technologies in the fight against gender inequalities. Without privacy and safety at the core of technology development, truly equal and future-ready systems cannot be achieved. However, a clear vision of what feminist resistant technology truly looks like is still missing. This is a complex question---one that no single author or perspective can fully answer alone. The aim of this paper is thus more modest: to explore what can be built now to prepare for future climate crises. From a feminist perspective, we know that climate change will exacerbate existing inequalities. We have already seen how attacks on women's rights, such as abortion access in the U.S. and elsewhere, intensify in times of instability.

\section{How: Critical approaches to design}

We discuss in this section few critical approaches to designs and how they
fail to address the question proposed in this paper which is how to build
resistance technologies. These approaches however remain valuable and we
are not arguing otherwise. In fact they should be used for future designs. Unfortunately these frameworks have been used to alternative systems such as car sharing or mediations delivery, systems that are considered alternatives.
Therefore we argue that they should be complemented with other approaches, for example
counter-factual history proposed by Eriksson et. al in order to imagine and prepare designs for times of polycrisis~\cite{eriksson2018meeting}.

\subsection{Collapse Informatics}

Collapse Informatics is a term that has been coined by Tomlinson et al. in~\cite{tomlinson2013collapse}.  It refers to the study and design of socio-technical systems in the context of anticipated societal and/or environmental decline, emphasizing resilience, adaptability, and sustainability under resource constraints. The authors explore how technology and design might support human well-being and argue for rethinking how we approach design in anticipation of \enquote{collapse} scenarios---such as climate change, energy descent, or economic instability---where assumptions of growth, abundance, and continuity may no longer hold. Additionally the authors criticize traditional practice theory arguing that it emphasizes practices that are habitual, unconscious, and reproduced routinely without much deliberate reflection instead we should integrate practice theory to prepare for societal collapse that requires more than reactive shifts during crises. Unlike traditional views that see practices as unconscious routines, the authors highlight the importance of consciously reflecting on and intentionally transforming behaviors to build resilience. This reflexive practice involves communities critically evaluating and redesigning their habits and technologies to proactively adapt to potential collapse conditions, making such intentional change a key focus for sustainable, future-oriented design.

\subsection{Sustainable Software Designs}

\subsubsection{Practical Perspectives on Sustainability in Software Design}

A notable contribution to the practical perspective on sustainability is
found in the paper \enquote{Requirements Engineering for Sustainability: An
Awareness Framework for Designing Software Systems for a Better Tomorrow.} This paper introduces a question-based framework aimed at raising awareness about the potential social, economic, and environmental impacts of software systems. Recognizing that software systems can have far-reaching consequences, the authors advocate for a paradigm shift in software design that maintains or improves sustainability within socio-technical systems.

The framework helps requirements engineers, who may lack the necessary knowledge and methodological support, to facilitate discussions about sustainability during the software development process. The authors conducted an evaluation study with four groups of computer science students, demonstrating that the framework encourages the consideration of sustainability effects and is adaptable to different types of systems. This represents an early step in fostering a paradigm shift in software engineering practices.

Additionally, the Sustainability Awareness Framework (SusAF), proposed by the Sustainability Alliance for Sustainability Design, is another significant tool. SusAF helps companies and other stakeholders to identify potential sustainability effects, both environmentally and socially, before designing software. As highlighted in the case study by Porras et al.~\cite{porras2021could}, this framework allows companies to enhance their awareness of the sustainability impacts of their technological products and services.

\subsubsection{Karlskrona Manifesto for Sustainability Design}

The Karlskrona Manifesto for Sustainability Design is a key document that advocates for the integration of sustainability as a core concern in the design of software systems. The manifesto stresses the profound impact software systems have on society and the environment and calls for responsible design practices that consider long-term consequences. The key principles of the manifesto are as follows:
\begin{itemize}
	\item Sustainability is systemic: Sustainability requires a systems thinking approach to understand the interconnections within and beyond the software system.
	\item Sustainability has multiple dimensions: These include environmental, social, economic, individual, and technical aspects, all of which must be considered during the design process.
	\item Sustainability transcends disciplines: Addressing sustainability challenges necessitates collaboration across various fields and perspectives.
	\item Sustainability applies to both the system and its broader contexts: Designers must consider the sustainability of the system itself and its impact on the wider environment and society.
	\item Sustainability requires long-term thinking: Design decisions should account for long-term effects, ensuring that current solutions do not compromise future needs.
\end{itemize}

\subsubsection{The Bits and B\"aume Conference 2022}

The Bits and B\"aume Conference 2022 presented a comprehensive set of over 60 political demands aimed at steering digitalization toward sustainability, social justice, and democratic governance. These demands are organized into five key areas:
\begin{enumerate}
	\item Digitalization within planetary boundaries: Digitalization must respect planetary limits by building climate-neutral infrastructure, reducing data flow, and supporting repairable, open devices.
	\item Global justice and regional empowerment: Local communities, including small-scale agriculture and Indigenous groups, must be involved to ensure global digital justice.
	\item Redistribution of technological power, democracy, and participation: The manifesto calls for the regulation of monopolies, banning exploitative tracking, and promoting open-source and public-interest models.
	\item Justice in digitalization, sustainable technology design, and social issues: Technology must be inclusive, peace-promoting, safe, accessible, and free from discrimination.
	\item Strengthening the common good through education, research, and public infrastructure: Promotes the strengthening of public infrastructure, education, and research to support democratic digitalization.
\end{enumerate}

\subsubsection{Manifesto for Energy-Aware Software}

In their article \enquote{A Manifesto for Energy-Aware Software,} Alcides Fonseca, Rick Kazman, and Patricia Lago discuss the growing energy consumption within the computing and communications sector, projected to account for 20\% of global energy use by 2025. They argue that energy considerations are often overlooked in the software engineering community and advocate for the development of energy-aware software systems designed to monitor and respond to energy usage patterns.

The manifesto introduces nine guiding principles for fostering energy awareness among stakeholders, emphasizing the need for a shift in mindset and practices to integrate energy considerations into the software development process. Public awareness, the authors argue, is crucial for the widespread adoption of energy-aware practices in the software industry.

We highlight some common themes across these political demands and manifestos. All of these voices agree on the importance of interdisciplinary work, the notion that sustainability transcends technical solutions and requires a holistic approach, and that designing systems necessitates a critical perspective from the outset. We fully support these demands and calls. However, we also note the vagueness in some of these demands, particularly when it comes to engineering practices. For instance, while we agree that sustainability has multiple dimensions (e.g., social, environmental), how can we incorporate these dimensions into a system's design or code? For example, how can we analyze the societal impact of a cryptographic primitive?
In the next section, we examine two main design frameworks, Design Justice and Just Sustainability, proposed in the literature. Similar to the manifestos and calls related to design, we highlight the common themes within these frameworks and explore how they can help us answer the critical question: \enquote{What should we design?}

\subsection{Design Frameworks}

\subsubsection{Design Justice}

In her book, Sasha Costanza-Chock highlights the importance of design principles such as affordance, bias, and open-source practices. She emphasizes that the skills of marginalized communities, which have traditionally been undervalued by capitalism, should be incorporated into design processes. 

Costanza-Chock introduces the concept of \enquote{Design Justice,} a framework that analyzes how the design of socio-technical systems influences the distribution of benefits and burdens among different groups. She proposes several design principles that can help create more inclusive systems, such as prioritizing the community's impact, seeing the designer as a facilitator rather than an expert, and sharing design knowledge and tools.

Design Justice does not prescribe a single optimal design approach but
instead offers a way to evaluate whether designs contribute to dismantling
or reproducing inequalities. It serves as both a framework for analysis and
a growing community of practice. Patricia Collins further expands on this,
using the concept of the \enquote{matrix of domination} (including white supremacy, capitalism, and settler colonialism) to explain how design principles can either reinforce or challenge societal structures. Patricia Hill Collins emphasize that in order one design principle should be to identify the communities of resistance, particularly focusing on the ways marginalized groups challenge systems of oppression. She highlights the importance of Black feminist thought as a source of resistance and empowerment, recognizing that communities of color and women of color develop unique strategies for resisting oppression. Collins' work emphasizes the interconnectedness of race, class, and gender in shaping experiences of power and domination, leading to the development of distinct forms of resistance.

\subsubsection{Just Sustainability}

As Christoph Becker discusses in his book, the concept of green
IT---replacing existing technologies with energy-efficient alternatives---has a
significant limitation: consumption rebounds. For instance, while
energy-efficient light bulbs reduce energy consumption per unit of light,
the total energy consumed often remains the same due to increased usage.
Becker argues that the core challenge of \enquote{just sustainability} is not technical, economic, or scientific, but rather ethical. The most pressing issue is determining how to protect the most vulnerable populations amidst the climate crisis.

Becker~\cite{becker2023insolvent} identifies several myths in the realm of computing that hinder progress towards sustainability:
\begin{enumerate*}[label=(\roman*)]
	\item Technology is value-neutral.
	\item The human is a flawed computer.
	\item Problems exist and can be solved.
	\item Design is problem-solving.
\end{enumerate*}
Becker concludes that computing is not yet ready for just sustainability, but
can become so through collaboration with \enquote{critical friends} who can guide the process of transforming IT into a force for good. Becker's work illustrates how computing has stagnated and emphasizes the importance of critical engagement to overcome this stagnation.
While these design frameworks are valuable, they tend to focus on the
\enquote{how} and the \enquote{why,} often neglecting the \enquote{what.}
For engineers, PhD students, and academics who are already attuned to the
ethical and societal challenges, the frameworks clarify \enquote{how} to design and build but frequently fall short of guiding what should actually be built. They address many important questions, but often leave unanswered the most urgent ones---particularly in times of crisis. This is, of course, a difficult challenge. As Eriksson et al. note, there are \enquote{limitations in our own thinking about, for example, future technical systems and in our ability to imagine futures that are characterized by various limitations.}

\subsection{The role of academic institutions}

Academics have long been at the forefront of highlighting the dangers of
climate change, and many argue that universities must match the demands
with sufficient action. Despite thousands of institutions declaring a
\enquote{climate emergency,} the higher education sector often continues
with business as usual, lacking the urgency needed to address the crisis
effectively. In their 2021 call, \enquote{From Publications to Public
Actions,} Gardner et al. argue that while universities promote
sustainability through research and teaching, these efforts alone are
insufficient~\cite{gardner_publications_2021}. They advocate for
universities to support academic advocacy and activism, including endorsing
non-violent civil disobedience. Similarly, the call \enquote{Preaching
Water While Drinking Wine} critiques universities for not aligning their operations with their climate commitments~\cite{borgermann_preaching_2022}. It calls for institutions to address all scopes of emissions and set science-based reduction targets, urging them to lead by example in the fight against climate change .

To truly lead in climate action, universities must integrate sustainability into every aspect of their operations. This includes reducing air travel, divesting from fossil fuels, and supporting climate activism among students and staff~\cite{capstick_civil_2022}.

While we support and share the concerns expressed in these calls for academic engagement with the climate crisis, we argue that the role of universities, academics, and students must go beyond advocacy. There should be a pressing need to actively produce knowledge, practices, and infrastructures that prepare society for living through polycrisis. In this context, scholars and institutions have an ethical responsibility to contribute not only to critique but also to the development of tangible systems to emerging global challenges.

\section{What: Resistance Technologies or Technologies during Crisis?}

In this section, we explore what kinds of technologies we should aim to design in anticipation of crises. Of course, we do not claim this question has an easy or definitive answer. Instead, we attempt to outline a set of features that may characterize \textit{resistance technologies}. We begin by examining technologies that have been deployed during crises, highlight their limitations, and explain why they often fall short of becoming sustainable resistance technologies.

\subsection{Technologies during Crisis}

During crises, technology has simultaneously been both actively rejected and heavily relied upon. For example, during the mass protests in Hong Kong, demonstrators dismantled surveillance infrastructure, destroying smart poles equipped with facial recognition, cutting electronic wiring, and using laser pointers to disrupt citywide surveillance systems~\cite{honkkong2019}.

At the same time, technology has long been a crucial asset in crisis response, enabling resource coordination, disaster prediction, and recovery efforts. For instance, the Singaporean government developed OneMap, an AI-based platform designed to anticipate future risks. Satellite-based sensors have been used to detect and monitor natural hazards. During the 2018 California wildfires, Geographic Information Systems (GIS) provided real-time identification of high-risk areas. Following the 2010 Haiti earthquake, the Ushahidi platform empowered citizens to report urgent needs and available resources, enhancing grassroots coordination.
~\cite{gillespie2007assessment, moloney2025routledge}.

In public health, technology has become increasingly critical. The COVID-19
pandemic saw the rise of contact-tracing applications used to monitor and
manage viral spread. Crisis-oriented technologies are not exclusive to
state-level initiatives; communities and individuals also develop their own
tools. For instance, during the Hong Kong protests, activists used
Bridgefy, a peer-to-peer communication app that relies on decentralized
short-range Bluetooth connections, during internet
shutdowns. In Gaza, Palestinians have leveraged eSIM technology, supported
by an Egyptian organization, to stay connected amid infrastructure
collapse.

While these examples demonstrate valuable efforts, several key shortcomings remain:
\begin{itemize}
  \item Lack of autonomy: In times of crisis, such as during the invasion of Ukraine, a lack of technological autonomy can be perilous. When Russia disrupted internet access, Ukraine relied heavily on Starlink for communication and logistics. However, this dependency has become a vulnerability, as Elon Musk has recently threatened to restrict access~\cite{pollet2025ukraine}.
  \item Significant privacy and security vulnerabilities: Bridgefy is an instant messaging app that operates over Bluetooth, enabling users to communicate without an internet connection. It gained popularity during the Hong Kong protests for this reason. However, the app suffers from serious security and privacy vulnerabilities, which can be easily exploited by attackers. For instance, an attacker can easily intercept and decrypt messages, impersonate users, and even track users' movements~\cite{albrecht2022breaking}. Similarly, many COVID-19 contact-tracing apps, while crucial in helping to curb the spread of the virus, also raised significant concerns due to weak security measures and privacy violations~\cite{ang2021covid}.
  \item Since October 2023, Palestinians in Gaza have faced repeated internet blackouts. In response, an Egyptian NGO, Connecting Humanity, led by the activist Mirna El Helbawi, launched the ConnectingGaza campaign. Through this initiative, individuals in Gaza contact the NGO, and El Helbawi sends them activation codes for eSIMs to help restore connectivity. While this is a great initiative---recognized with an award from the Electronic Frontier Foundation---it highlights serious scalability issues. Relying on a single individual or group to manually distribute eSIMs is not viable in the long term, especially during widespread or prolonged crises~\cite{gaza2023esim}.
\end{itemize}
We note that the above shortcomings are not exhaustive. While we have highlighted some of the most pressing issues, others certainly exist. We want to emphasize two key points. First, these issues are deeply interconnected. For example, while the primary challenge of the eSIM project in Gaza is scalability, it also raises concerns related to security and autonomy. Similarly, Starlink's main issue lies in its reliance on a single centralized entity (effectively one individual, Elon Musk) who can make decisions on behalf of the entire Ukrainian population; it also lacks robust privacy protections. Second, these shortcomings are largely a result of inadequate preparation before crises occur. In emergency situations, we often resort to rapid, improvised solutions that overlook critical ethical, social, and technical considerations. It is only later that the long-term consequences of these designs become apparent. Unsurprisingly, none of these systems underwent a proper life cycle assessment or were evaluated through the lens of just sustainable design.

There is a critical distinction between technologies deployed as reactive measures (e.g., laser pointers during protests) and those that are proactively developed to address foreseeable challenges (e.g., decentralized communication systems).

Building reliable, secure, privacy-respecting, and sustainable technologies requires long-term effort and collective collaboration. Unfortunately, such development is often neglected until a crisis emerges, resulting in rushed, brittle, and unsustainable solutions. Later, the absence of privacy or sustainability is rationalized as a necessary compromise due to the urgency of the situation. But should it be?

The sustainability community often refers to the Brundtland Report's widely cited definition of sustainable development~\cite{brundtland_our_1987}:

\begin{quote} Sustainable development is development that meets the needs of the present without compromising the ability of future generations to meet their own needs. \end{quote}

Yet, the goal of sustainability may go beyond merely avoiding harm. It is also our responsibility to design and build systems that help people cope with the crises produced by current and past generations.
To address this, we turn to the literature on privacy, recognizing the significant similarities, challenges, and future visions shared by the sustainability and privacy communities.

\subsection{Lessons learned from Privacy Engineering}

\enquote{Data is the new oil} is a term coined in 2006 by Clive Humby, a mathematics professor who has been at the forefront of innovation in consumer data. He draws an analogy between oil and data, emphasizing their critical role in driving business and innovation. This vision of data as the \enquote{new oil} has led to the extensive collection of personal data, resulting in numerous privacy violations and the rise of surveillance capitalism~\cite{zuboff2023age}. As Bart Preneel points out, \enquote{If data is the new oil, data mining yields the rocket fuel}, suggesting that just as the oil industry has caused numerous climate disasters, data mining is responsible for many social disasters with surveillance being one of the most harmful~\cite{preneel2016future}.

In response, an entire body of scholarship emerged, and the term Privacy by Design was coined by Ann Cavoukian in 2009~\cite{cavoukian2009privacy}, outlining principles such as data minimization and privacy by default to guide the development of privacy-enhancing systems.
However, as G\"urses et al. argue,
the term Privacy by Design has struggled to meaningfully impact the design
of privacy-enhancing technologies. One key issue is that its definition and
principles were often too vague and ambiguous for engineers to design and
\enquote{to be translated into the engineering practice}~\cite{gurses2011engineering}.  For example one of the principles proposed by Cavoukian is Privacy Embedded into Design which only communicated that privacy is important in the design but unclear what and how to be implemented.

To address this disconnect, G\"urses et al. have systematized the knowledge of privacy engineering and argues that in system design any form of data collection should be viewed as a potential privacy loss, and thus the trade-offs involved must be critically assessed. A central challenge is to avoid function creep---the unintended re purposing of a function beyond its original purpose. For example, a parking system feature intended only to measure how long a user use the parking (to calculate fees) could be repurposed to infer sensitive behavioral patterns, such as how frequently someone visits a hospital or which department they frequent. The goal should be to build essential system features (e.g. functionality) without enabling such creep~\cite{gurses2011engineering}

Jaap-Henk Hoepman further contributed to this discourse by classifying the the privacy by design approaches into two categories: soft and hard. Soft approach rely on legal and organizational control mechanisms, assuming that if a privacy violation occurs, consequences will be enforced through regulatory or institutional channels. In contrast, hard approaches aim to achieve privacy by three main principles:
\begin{enumerate*}[label=(\roman*)]
  \item avoiding a single point of failure,
  \item considering that any exposure of data is a lost of privacy and
  \item increasing trust in the privacy solution itself not in the control mechanism that will ensure privacy.
\end{enumerate*}
This approach rely on designing systems in such a way that privacy violations are technically prevented from the outset~\cite{hoepman2021privacy}.
Like sustainability, privacy is complex and holistic with no precise
definition or properties. Privacy by design researchers have been
advocating for inter-disciplinary work, strategies and principles that help
build privacy-enhancing technologies for years. This effort led to the book
\enquote{Privacy is hard and seven other myths: Achieving privacy through careful
design} where Hoepman identified 8 strategies with various different technical and legal tactics that lead to engineering and building privacy technologies.

This should not be misunderstood as suggesting that engineers---and
only engineers---should be responsible for designing privacy-enhancing
systems. On the contrary, G\"urses emphasizes that \enquote{privacy by design
is more than just a matter of technological design} highlighting the
long-standing consensus within the global privacy community on the
importance of interdisciplinary collaboration. However, the privacy community has also emphasized that privacy technologies require principles and strategies that are clearly defined and translatable into implementation~\cite{gurses2011engineering}.

We identify similar challenges, principles as well as goal from both communities: the privacy community and the sustainability community. The core principles advocated by the sustainability
community, such as interdisciplinary collaboration, evaluation of
resources and infrastructures, thinking critically about the design from the beginning of the life cycle of software are also crucial for designing privacy tech. This is highlighted by M\"uhlberg in~\cite{muehlberg_2022_sustaining_sec} where he argues for appropriating the notion of critical refusal in order to have not only sustainable software but also security and safety.
Therefore by combining principles from just and sustainable design with those from Privacy by Design---particularly its engineering dimension---we summarize our findings in the form of key guiding questions. Our hope is that these questions will support both engineers and the broader \textit{Computing within Limits} community.

\subsection{Resistance Technologies}

How, then, can we define \textit{resistance technology}? Resistance, as a political term, is interpreted differently across communities. However, there is a shared understanding that it entails a refusal to be subjugated and a commitment to re-imagining how humans live on this planet\footnote{\url{https://theecologist.org/2018/jun/05/how-environmental-justice-movement-transforms-our-world}}.

Combining lessons from Privacy by design and from the Design principles proposed by the sustainability community we propose that \textit{resistance technologies} should reflect on the following questions when it comes to engineering and designs:

\begin{itemize}
	\item Reflect on past crises: Anticipate the needs of future crisis
to assist current and future generations facing a (poly-)crisis.
	\item Cooperation over competition: Operate outside of market logic, prioritizing cooperation over competition.
	\item Solidarity: Unlike eco-survivalism, which emphasizes individual preparation for collapse~\cite{katz2022preparing}, we argue that community-focused responses should be central to resistance technologies, as they mitigate isolation and self-interest while fostering collective resilience.
	\item Minimize dependence on resources provided by large tech corporations (e.g., AWS), especially those that can be withdrawn at any time. Sovereignty should be a core design goal.
	\item No centralization: Avoid reliance on centrally provided or controlled resources that may become unavailable or untrustworthy during crises.
	\item Critical refusal: M\"uhlberg has argued that some ICT features such as security and safety are more robust when rooted in critical refusal~\cite{muehlberg_2022_sustaining_sec}, a term previously coined in~\cite{cifor2019feminist}.
	\item Avoid function creep: particularly in systems that centralize data under the pretext of utility but pave the way for these same functions to be used differently (data collection).
\end{itemize}

Finally, we conclude that Privacy-Enhancing Technologies are essential technologies that must be designed and implemented in response to the crises driven by climate change. As discussed in Section 2, privacy is crucial for the Global South, enabling gender equality, and during wars and armed conflicts.

\section{Discussion and Future Work}

Technological development under capitalism has largely been dictated by the global market. The private sector and, to a large extent, academia have prioritized the production of technologies that align with the logic of market competition and profit maximization. This is not to say that there are no communities that have been active in resisting these technologies; the sustainability and privacy communities (and probably others) have been producing guidelines, design principles and even software that challenge this status quo.

However, much of this work has focused on creating alternatives that provide the same functionalities of capitalist technologies, albeit with greater emphasis on values such as privacy and sustainability. Unfortunately most of these projects have been buried in the graveyard of the long list of projects that could not compete in the market. 

We argue that what is needed in a move towards just sustainability is a paradigm shift: instead of trying to
create market-viable alternatives, we must focus on producing Resistance
Technologies. The goal of these technologies should not be replacing or being an alternative to current technologies, rather to be useful during times of crisis. For
example, rather than developing a \enquote{sustainable AI for the Global
South} as a concept that will likely increase technological dependency and
inequality\footnote{For different perspectives cf., for example, Tony Blair Institute for Global Change:
\enquote{Powering AI in the Global South},
\url{https://institute.global/insights/climate-and-energy/powering-ai-in-the-global-south};
World Economic Forum: \enquote{The 'AI divide' between the Global North and
the Global South},
\url{https://www.weforum.org/stories/2023/01/davos23-ai-divide-global-north-global-south/};
Rachel Adams: \enquote{AI Is Bad News for the Global South},
\url{https://foreignpolicy.com/2024/12/17/ai-global-south-inequality/}.} we might instead focus on technologies that facilitate the
distribution of resources, developed as a common good and with a focus on sufficiency and resilience,
that are designed to facilitate aid during climate-related disasters.
It is important to understand such technological development in the context
of available resources and systemic sustainability of human activity within
planetary boundaries.

O'Neill et al.~\cite{oneill_good_2018} ask \enquote{what  level of biophysical resource use
is associated with meeting people's basic needs, and can this level of resource use be extended to all
people without exceeding critical planetary boundaries.} They conclude that
a fundamental restructuring of distribution systems paired with increased
efficiency can help societies to move much closer to a safe and just space,
if we focus development on notions of sufficiency and degrowth.
In~\cite{millward-hopkins_providing_2020} Millward-Hopkins et al. argue
further that the help of advanced technologies across all sectors, paired with radical
demand-side changes to reduce consumption, can allow us to reach this safe
and just space for everyone. Thus, lifestyle choices that lead towards
strong sustainability do not need to come from perspectives of scarcity but
can be designed for
sufficiency or even abundance. These lifestyles can led to globalized
societies and actually improve living conditions
for a majority of people through redistribution of resources and power.
We see specifically the latter, redistribution of power through equal access and
participatory approaches to governance (cf.~\cite{bouzguenda2019towards,
hawkins2012sustainable}) as an important driver for the development of
sustainable resistance technologies that are perceived as a \enquote{more}
in terms of freedom and possibilities rather than an inconvenience and a
restriction on personal consumption.
Our notion of \emph{Resistance Technologies} aim
to define some design objectives for the advanced technologies of a
sufficient and resilient future that enable decent living for all while
promoting preparedness for the worst.

In this paper, we argued that privacy---as in anti-surveillance---is a core value of these Resistance Technologies. We ask the question to the Computing within Limits community: what other core values should guide the design and development of technologies that aim not to compete, but to sustain and protect us in a world facing ecological and economic crises?

Like privacy, resistance technology is unlikely to be defined by a single, universally accepted concept. It is not a plug-and-play solution that automatically makes systems sustainable, private, and resistant. Instead, we identify a list of features and reflective questions that can help us design future designs. We invite the \textit{Computing within Limits} community to engage with, critique, and extend this discussion towards further developing Becker's \enquote{critical friends}~\cite{becker2023insolvent} as guiding principles for \emph{Resistance Technology}.

\begin{acks}
We gratefully acknowledge the Brussels-Capital Region -- Innoviris for
financial support under grant numbers 2024-RPF-2 MImPG and 2024-RPF-4 SDM, and the
CyberExcellence programme of the Walloon Region, Belgium (grant
2110186).
\hyphenation{Innoviris}
\end{acks}

\bibliographystyle{ACM-Reference-Format}
\balance
\bibliography{bibfile}

\end{document}